\documentclass[11pt,preprint]{aastex}
\usepackage{natbib}

\shorttitle{BIMA/OVRO observations of comet C/1999 S4}
\shortauthors{Hogerheijde et al.}

\begin{document}

\title{Combined BIMA and OVRO observations of comet C/1999 S4 (LINEAR)}

\author{
Michiel R. Hogerheijde\altaffilmark{1,2,3}, 
Imke de Pater\altaffilmark{1}, 
Melvyn Wright\altaffilmark{1},
J. R. Forster\altaffilmark{4}, 
L. E. Snyder\altaffilmark{5},
A. Remijan\altaffilmark{5}, 
L. M. Woodney\altaffilmark{6},
M. F. A'Hearn\altaffilmark{7}, 
Patrick Palmer\altaffilmark{8},
Y.-J. Kuan\altaffilmark{9,10}, 
H.-C. Huang\altaffilmark{9,10}, 
Geoffrey A. Blake\altaffilmark{11,12}, 
Chunhua Qi\altaffilmark{11,13}, 
Jacqueline Kessler\altaffilmark{12}, 
S.-Y. Liu\altaffilmark{10}}

\altaffiltext{1}{Astronomy Department, University of California, 601
Cambell Hall \# 3411, Berkeley, CA 94720-3411, USA}
\altaffiltext{2}{Steward Observatory, The University
of Arizona, 933 N. Cherry Ave, Tucson, AZ 85721, USA}
\altaffiltext{3}{Current address: Sterrewacht Leiden, P.O. Box 9513,
2300 RA, Leiden, The Netherlands}
\altaffiltext{4}{Hat Creek Radio Observatory, 42231 Bidwell Rd, Hat
Creek, CA 96040, USA}
\altaffiltext{5}{Department of Astronomy, University of Illinois, 1002
W. Green Street, Urbana, IL 61801, USA}
\altaffiltext{6}{Lowell Observatory, 1400 Mars Hill Rd, Flagstaff, AZ
86001, USA}
\altaffiltext{7}{Astronomy Department, University of Maryland, College
Park, MD 20742, USA}
\altaffiltext{8}{Department of Astronomy and Astrophysics, University
of Chicago, Chicago, IL 60637, USA}
\altaffiltext{9}{Department of Earth Sciences, National Taiwan Normal
University}
\altaffiltext{10}{Academia Sinica, Institute of Astronomy and
Astrophysics, P.O. Box 23-141, Taipei 106, Taiwan, ROC}
\altaffiltext{11}{Division of Geological and Planetary Sciences,
California Institute of Technology 150-21, Pasadena, CA 91125, USA}
\altaffiltext{12}{Division of Chemistry and Chemical Engineering,
California Institute of Technology, Pasadena, CA 91125, USA}
\altaffiltext{13}{Harvard-Smithsonian Center for Astrophysics, 60
Garden Street, Cambridge, MA 02138, USA}

\email{michiel@strw.leidenuniv.nl}


\begin{abstract}
We present results from an observing campaign of the molecular content
of the coma of comet C/1999~S4 (LINEAR) carried out jointly with the
millimeter-arrays of the Berkeley-Illinois-Maryland Association (BIMA)
and the Owens Valley Radio Observatory (OVRO). Using the BIMA array in
autocorrelation (`single-dish') mode, we detected weak HCN $J$=1--0
emission from comet C/1999~S4 (LINEAR) at $14 \pm 4$~mK~kms averaged
over the $143''$ beam. The three days over which emission was
detected, 2000 July 21.9--24.2, immediately precede the reported full
breakup of the nucleus of this comet. During this same period, we find
an upper limit for HCN 1--0 of 144~mJy~beam$^{-1}$~km~s$^{-1}$
(203~mK~kms) in the $9''\times 12''$ synthesized beam of combined
observations of BIMA and OVRO in cross-correlation (`imaging')
mode. Together with reported values of HCN 1--0 emission in the $28''$
IRAM~30-meter beam, our data probe the spatial distribution of the HCN
emission from radii of 1300 to 19,000~km. Using literature results of
HCN excitation in cometary comae, we find that the relative line
fluxes in the $12''\times 9''$, $28''$ and $143''$ beams are
consistent with expectations for a nuclear source of HCN and expansion
of the volatile gases and evaporating icy grains following a Haser model.
\end{abstract}



\section{Introduction\label{s:intro}}

The volatile component of comets provides an excellent tracer of the
chemical composition present during the formation of the Solar System
\citep{mumma:ppiii, irvine:iau197, irvine:ppiv}. Recent years have
seen remarkable progress in this area with the appearances of two
especially active comets, Hyakutake (C/1996 B2) and Hale-Bopp (C/1995
O1), that passed close to Earth. Observations at (sub) millimeter
wavelengths of these spectacular comets, in addition to numerous less
prominent but no less interesting comets over the last decade,
indicate that the composition of comets is heterogeneous. This holds
not only from comet to comet (e.g., \citealt{ahearn:ensemble}), but
also within the same comet as was shown with
millimeter-interferometers that resolve the emission from the coma
(e.g., \citealt{gab:halebopp}). The latter work illustrated the
information that high spatial resolution observations can offer about
cometary composition and evaporation processes.

In the Summer of 2000, the comet C/1999 S4 (comet LINEAR from here on)
passed close to the Sun (0.765~AU) and at 0.374~AU to the Earth
\citep{durig:linear}. Its orbit identified it as a dynamically young
comet, and it showed a high level of activity at large heliocentric
distance. As is not uncommon in dynamically young comets, the increase
in its activity lagged significantly behind what was expected after
its discovery, reaching only a medium level of activity at perihelion
passage. It generated some excitement when, at closest approach to the
Sun, the nucleus of the comet disintegrated and the residual material
faded away in a few days, first reported by
\citet{kidger:iauc7467}. The comet's characteristics and its breakup
are discussed in detail by (\citealt{bockelee:linear, farnham:linear,
lisse:linear, makinen:linear, mumma:linear, weaver:linear,
bonev:linear, altenhoff:linear, tozzi:linear, schulz:linear,
cochran:linear, jockers:linear, rodriguez:linear, kidger:linear}, and
\citealt{magee-sauer:linear}). Of most relevance to our discussion
here are: the conclusion that the comet's nucleus may have broken up
already several weeks before it fully disintegrated in late July (July
23--24); the molecular-line observations reported by Bockel\'ee-Morvan
et al. including the HCN-production rate; and the report by
\citet{magee-sauer:linear} of infrared HCN line observations.

This Paper describes the results of an observing campaign of the
volatile composition of comet LINEAR, carried out jointly by the
millimeter arrays at Hat Creek (operated by the
Berkeley-Illinois-Maryland Association) and Owens Valley (operated by
the California Institute of Technology), both in California (\S
\ref{s:obs}). We present a weak detection of HCN $J$=1--0 in
auto-correlation mode and upper limits on the emission in
cross-correlation mode, and place limits on the spatial extent of the
emission (\S \ref{s:results}). Using models of HCN excitation in
comets from \citet{crovisier:rovibspectra} (see also
\citealt{bockelee:comethcn}) we investigate whether the spatial extent
matches expectations (\S \ref{s:model}). Our conclusions are
summarized in \S \ref{s:conclusions}. The Appendix discusses the
methods used to combine data from the two millimeter arrays.

\section{Observations\label{s:obs}}

Observations of comet LINEAR were obtained between 2000 July 17 and
2000 July 27 at the millimeter interferometers of the
Berkeley-Illinois-Maryland Association (BIMA)\footnote{The BIMA Array
is operated by the Berkeley-Illinois-Maryland Association under
funding from the National Science Foundation.} at Hat Creek,
California ($121^\circ\, 28'\, 18{\farcs}49$ West, $40^\circ\, 49'
2{\farcs}5$ North; altitude 1043~m), and the Caltech Millimeter Array
at the Owens Valley Radio Observatory (OVRO)\footnote{The Owens Valley
Millimeter Array is operated by the California Institute of Technology
under funding from the U.S.\ National Science Foundation
(\#AST96--13717).} near Big Pine, California ($118^\circ\, 16'\,
55{\farcs}92$ West, $37^\circ\, 14'\, 2{\farcs}04$ North; altitude
1222~m). The six 10-meter antennas of the OVRO array were in `compact'
configuration, the ten 6-meter antennas of the BIMA array in `C'
configuration: both configurations result in comparable projected
baselines: 2--20 k$\lambda$ at BIMA and 3--15 k$\lambda$ at OVRO.
Since these were summer-time observations, when atmospheric stability
can be poor for millimeter-interferometry, these compact
configurations were adopted to avoid decorrelation by atmospheric
phase fluctuations on long baselines.

OVRO observed the comet in crosscorrelation mode from rise in the
early morning local time until set in the late afternoon local
time. BIMA operated in crosscorrelation mode from rise until
mid-morning local time. After mid-morning the phase stability at Hat
Creek deteriorated significantly during the observing period, and the
observing mode was switched to autocorrelation. This observing mode1
does not suffer from phase-decorrelation but has a large beam-dilution
in the $143''$ full-width at half-maximum (FWHM) primary beam of the
six-meter antennas. The phase stability at Owens Valley, approximately
500~km south-east of Hat Creek and on the opposite side of the
mountain range of the Sierra Nevada, was significantly better compared
to Hat Creek, and successful crosscorrelation data were obtained
throughout each day. In addition, the digital correlator at OVRO
cannot operate in autocorrelation mode, which fortunately was not a
limitation.

Receivers and digital correlators of both instruments were configured
to record the lines of HCN $J$=1--0 (88.6318470 GHz) and HCO$^+$
$J$=1--0 (89.1885230 GHz). At OVRO each of the HCN and HCO$^+$ line
frequencies were recorded in a single band of 64 channels with a
resolution of 125~kHz (0.42 km~s$^{-1}$).  At BIMA the lines were
recorded in two bands per line, centered on the line frequency, with
128 and 256 channels each and with respective channels widths of
391~and 98~kHz (1.3 and 0.33~km~s$^{-1}$). The autocorrelation mode of
the BIMA correlator offers twice the number of channels at half the
width for each of the bands, giving velocity resolutions of 0.65 and
0.17~km~s$^{-1}$.

Great care was taken to ensure that both arrays were using the same
comet ephemeris at all times. On each of the observing days the most
recent orbit solution from the Jet Propulsion Laboratory Solar System
Dynamics Group\footnote{Available at {\tt
http://ssd.jpl.nasa.gov/horizons.html}} was used (Table
\ref{t:obs}). Changes of the predicted comet's position between
subsequent orbit solutions were less than a few arcseconds, smaller
than the synthesized beam of $\sim 12''$ and much smaller than the
primary beam of $143''$ for the BIMA autocorrelation observations. The
small day-to-day uncertainty in the comet's position therefore has no
appreciable effect on the data. Because OVRO and BIMA use different
coordinate systems (B1950.0 vs. J2000, respectively) and use,
respectively, geocentric and topocentric representations for the
coordinates of Solar-System objects, much effort was spent to ensure
that both instruments actually pointed at the same position on the
sky. Note that the parallax of the comet at a geocentric distance of
0.3~AU amounts to $\sim 2''$ over the distance between the two
arrays. The topocentric velocity of the comet was also Doppler tracked
at both sites, keeping the cometocentric velocity centered at
0~km~s$^{-1}$ in the spectral bands.

The complex gains of crosscorrelation data were calibrated separately
for the BIMA and OVRO data, although the same flux and phase
calibrators were used for both data. Depending on the position on the
sky of the comet, the quasars 0716+714, 0923+392, and 1156+295 were
used, with respective fluxes of 3.0, 1.2, and 1.2 Jy determined from
contemporaneous measurements of planets at OVRO (Table \ref{t:obs}).
Using the same calibrator at each observatory ensures that the flux
scaling of both data sets is identical. The OVRO data were calibrated
using the `MMA' package specific to Owens Valley
\citep{scoville:database}; the BIMA data were calibrated using the
`MIRIAD' software package \citep{sault:miriad}. The OVRO data,
originally stored in FITS format were subsequently converted to MIRIAD
format, and both crosscorrelation data sets were added together after
regridding to a common spectral resolution. The Appendix describes the
data-combination procedure in more detail, including a test case
validating the process. The combined data were then inverted and
imaged using the MIRIAD software package.

The autocorrelation observations were corrected for atmospheric
opacity and fluctuations by switching to an `off' position at $+20'$
in Right Ascension every 30 seconds. The data were recorded on the
main-beam brightness temperature scale in K; subsequent observations
of the Galactic SiO maser R~Leo in cross- and autocorrelation mode
confirmed the flux scaling.  A number of bad channels were flagged due
to spikes, and second- to fourth-order polynomial baselines were
subtracted from the autocorrelation spectra from the individual
antennas. Finally the spectra from the individual antennas were added
to yield a single autocorrelation spectrum for each of the spectral
bands. All processing of the autocorrelation data was carried out
within the MIRIAD software package.

\section{Results\label{s:results}}

\subsection{Autocorrelation spectra\label{s:auto}}

The autocorrelation spectra of HCN 1--0 (Fig. \ref{f:auto}) and
HCO$^+$ 1--0 were inspected using the average data of single,
individual days and averaging over two, three, and six consecutive
days. In HCN a weak signal was detected in two- and three-day averages
between 2000 July 21.85 and 2000 July 24.16. These are the three days
leading up to the disruption of the nucleus between 2000 July 24.9 and
25.9 \citep{kidger:iauc7467} and coincide with a reported brightening
at visible wavelengths, likely indicating a minor outburst event. The
three-day average main-beam brightness temperature of HCN 1--0 is
$14\pm 4$~mK~km~s$^{-1}$ in the $143''$ FWHM beam of the six-meter
dishes. The emission is centered around a velocity of
$-1$~km~s$^{-1}$, close to the expected velocity of the comet
(nominally at zero), and has a width of $\sim 2$~km~s$^{-1}$ FWHM. The
latter value is uncertain because of the low signal-to-noise of the
emission; the hyperfine components at +4 and $-7$~km~s$^{-1}$ were
likewise below the noise level. The reported intensity is a total for
all three components, obtained by multiplying the integrated intensity
of the single detected component by the expected relative intensities
in thermodynamic equilibrium: $0.2+1.0+0.6=1.8$. Although the overall
signal-to-noise of the detection is low, the three-day averages
leading up to (2000 July 21.85--24.16; $14\pm 4$~mK~km~s$^{-1}$) and
following (2000 July 24.95--27.17; $<10$ mK~km~s$^{-1}$ 2$\sigma$
upper limit) the comet's final breakup around 2000 July 23--24 clearly
show a dearth of the HCN 1-0 emission. The decrease in HCN $J$=3--2
emission observed with the IRAM 30-meter telescope by
\citet{bockelee:linear} after 2000 July 23.8 supports our detected
fall-off of HCN. No emission was detected in any of the HCO$^+$
spectra, to a 2$\sigma$ upper limit of 4~mK~km~s$^{-1}$ in three-day
averages.

\subsection{Crosscorrelation images\label{s:cross}}

The crosscorrelation data were inspected in a similar way as the
autocorrelation data. The data were integrated over a single day, over
multiple consecutive days, and over 4-hour time intervals to search
for transient features. This was done for the data from the two arrays
individually as well as for the combined data set. In all cases the
emission was integrated over 2 km~s$^{-1}$ around a cometocentric
velocity that was varied from $-10$ to $+10$~km~s$^{-1}$ to account
for possible small tracking offsets in the comet's velocity. The
reported flux (limits) refer to the sum of the hyperfine components,
added with the appropriate weights (1:5:3).

None of the images obtained in this way revealed any positive
detection of HCN 1--0 or HCO$^+$ 1--0 emission. Figure \ref{f:cross}
illustrates this for HCN 1--0 at an adopted cometocentric velocity of
0~km~s$^{-1}$ and including all OVRO and BIMA data. Using the
three-day average over 2000 July 21.85--24.16, over which emission was
detected in autocorrelation mode, an upper limit to the HCN 1--0
emission is found of 144~mJy~beam$^{-1}$~km~s$^{-1}$ (2$\sigma$) in
the $12''\times 9''$ synthesized (naturally weighted) beam,
corresponding to 203~mK~km~s$^{-1}$ . The corresponding values for
HCO$^+$ 1--0 are smaller, at 80~mJy~beam$^{-1}$~km~s$^{-1}$ and
117~mK~km~s$^{-1}$, because HCO$^+$ has no hyper-fine splitting to sum
over.

\section{Comparison to model calculations\label{s:model}}

The previous section showed that HCN 1--0 emission was detected, at
low signal-to-noise, in autocorrelation mode but not in
crosscorrelation mode. If the emission is unresolved ($<10''$), the
beam-averaged value of $14\pm 4$~mK~km~s$^{-1}$ detected in the
$143''$ primary beam corresponds to an intensity of $2.7\pm
0.8$~K~km~s$^{-1}$ in the $12''\times 9''$ synthesized beam of the
crosscorrelation data.  Since we obtained an upper limit of
203~mK~km~s$^{-1}$, it follows that the emission is significantly
resolved out in the interferometer beam. Alternative explanations for
reduced flux in the crosscorrelation data are decorrelation due to
uncorrected phase variations and positional uncertainties of the comet
larger than the synthesized beam. These processes are unlikely to
affect the data: the measured decorrelation on the phase calibrator
is typically less than 20\%, and, as mentioned in \S \ref{s:obs}, the
position of the nucleus was known to within a few arcseconds from day
to day, which is much smaller than the synthesized beam.

Extended emission is expected from molecules in the expanding
coma. For a parent species like HCN, the density in the coma follows a
Haser distribution,
\begin{equation}
n(r) = { Q \over {4\pi V_0 r^2}} \,
 \exp \Bigl ( - {{r-r_{\rm n}\over{r_\lambda}}} \Bigr),
 \label{e:haser}
\end{equation}
where $n$ is the number density of molecules, $Q$ the production rate
of that molecule by the nucleus, $V_0$ is the expansion velocity in
the coma, $r_{\rm n}$ is the size of the nucleus, and $r_\lambda$ is
characteristic photodissociation scale length. \citet{bockelee:linear}
quote $Q$(HCN)=$3\times 10^{25}$~s$^{-1}$; we adopt
$V_0=0.9$~km~s$^{-1}$ and $r_\lambda = V_0 / \beta = 3.7\times
10^4$~km, where $\beta=2.45\times 10^{-5}$~s$^{-1}$ is the HCN
photodissociation rate for the Solar spectrum at the time of the
observations (Bockel\'ee-Morvan 2001, private communication).

Is a distribution of HCN following eq. (\ref{e:haser}) consistent with
our detection in the $143''$ autocorrelation beam and our upper limit
in the $12''\times 9''$ crosscorrelation beam? Or is an additional
source of HCN required, distributed over $>10''$, e.g., evaporating
icy fragments that have broken off the nucleus? Fragments exceeding
$\sim 1$~m, non-uniformly illuminated by the Sun, are expected to
undergo non-gravitational forces and accelerate away from the inner
($\sim 10''$) coma (e.g., \citealt{desvoivres:nongrav}). Our data
uniquely probe the spatial distribution of the HCN 1--0
emission. Bockel\'ee-Morvan (2001, private communication) report HCN
1--0 emission in the $28''$ IRAM~30-meter beam of $93\pm
13$~mK~km~s$^{-1}$, averaged over 2000 July 21--23; we include this
data point in our analysis. We do not consider the HCN 3--2 emission
reported by \citet{bockelee:linear} here, because we limit our
discussion to the 1--0 transition.

The emission distribution of HCN 1--0 depends on the spatial
distribution of the molecules following eq.~(\ref{e:haser}) and on
their excitation as function of position. The latter is given by the
balance of (de-) excitation through collisions with water and
electrons, excitation into vibrationally excited levels through
absorption of Solar infrared photons, and de-excitation through
spontaneous emission. \citet{bockelee:comethcn} and
\citet{crovisier:rovibspectra} discuss the excitation of HCN in comets
in detail. Instead of carrying out a full excitation calculation, we
will use the HCN level populations modeled by
\citet{crovisier:rovibspectra}: this paper presents three cases: water
production rate $Q$(H$_2$O)=$2\times 10^{29}$~s$^{-1}$ and
heliocentric distance $r_h=1.0$~AU (model 1); $Q$(H$_2$O)=$2\times
10^{28}$~s$^{-1}$ and $r_h=1.6$~AU (model 2); and $Q$(H$_2$O)=$1\times
10^{30}$~s$^{-1}$ and $r_h=0.8$~AU (model 3). None of the parameters
of these models are exactly identical to those for comet LINEAR
($Q$(H$_2$O)=$3\times 10^{28}$~s$^{-1}$, \citealt{bockelee:linear};
and $r_h=0.76$~AU), but we show that especially the \emph{relative}
line intensities in the $12''\times 9''$, $28''$, and $143''$ beams
are comparatively insensitive to the model details and are largely
determined by the underlying column density distribution. Although the
exact values of the of the relative line intensities in the threee
beams depend on the details of the excitation distribution in the coma
that results from the balance between fluorescent and collisional
processes, values within the same range are found in the extreme cases
were either of these processes dominate. The models described above
can therefore be used to test if our data agree with a Haser
distribution for the column density.

Using a Haser distribution following eq.~(\ref{e:haser}) with the
parameters listed above, and the $J$=0 and $J$=1 HCN level populations
of the three models from \citet{crovisier:rovibspectra}, we calculate
the expected HCN 1--0 emission distributions at a distance of 0.374~AU
from Earth. We convolve the emission distributions with Gaussians of
$143''$ and $28''$ FWHM to get the main-beam intensities in these
beams. For the crosscorrelation data, we construct synthetic
visibilities based on the predicted emission distribution and the
actual antenna positions of the data. This explicitly takes into
account the spatial filtering of the interferometers and their
insensitivity to large-scale structure, as well as the data
combination procedure outlined in the Appendix. We process the synthetic
visibilities in the same way as the data to obtained
integrated-intensity images. We find that no more than 10--30\% of the
flux is resolved out by the interferometer.

Table~\ref{t:results} lists the observational results in the $143''$,
$28''$, and $12''\times 9''$ beams, alongside the predicted
intensities in these beams for the three models. In terms of absolute
intensities the observed values are within a factor of two of the
model results, indicating a general agreement between the models and
the observations. Table~\ref{t:results} also lists the observed and
modeled intensities normalized to the value in the $28''$ beam:
$F_{\rm auto}/F_{28}$ and $F_{\rm cross}/F_{28}$. For the three
models, $F_{\rm auto}/F_{28}$ ranges from 0.21 to 0.28. Our observed
value, $0.15\pm 0.05$, is consistent with the models at the $\sim
1\sigma$ level. Given the possibility of calibration differences
between the IRAM~30-meter data and the BIMA data, we conclude that
this indicates that the line emission over $143''$ is consistent with
the expectation of a Haser distribution when compared to a $28''$
region.  \citet{bockelee:linear} estimate that only fragments with
sizes $\gtrsim 1$~m are large enough to survive outside the inner few
thousand km of the coma. Our $143''$ result shows that such fragments,
if they exist, do not carry a significant contribution of the volatile
material ($\lesssim 50$\%). This is consistent with the results of
\citet{magee-sauer:linear}, who report that spatially resolved
infrared HCN line observations are consistent with a nuclear source,
and of \citet{bockelee:linear}, who conclude that the evolution of the
HCN production rate suggests that HCN is mainly released by icy
fragments of small size. 

The ratios $F_{\rm cross}/F_{28}$ of the three models show a larger
range, 1.57--3.65, compared to $F_{\rm auto}/F_{28}$. This indicates
that on these smaller scales, excitation differences exist depending
on the heliocentric distance and evaporation rate (= density of
collision partners). Our non-detection of emission in the
crosscorrelation beam yields a strict upper limit $F_{\rm
cross}/F_{28} < 2.5$. This is consistent with the range predicted by
the models, especially when also taking into account possible
decorrelation of our crosscorrelation data up to 20\% (see above) and,
again, possible calibration differences between the IRAM~30-meter and
OVRO+BIMA data. In other words, the non-detection in our combined
crosscorrelation data is expected for a Haser-model HCN
distribution. This indicates, that to obtain high-spatial resolution
(sub) millimeter observations of moderately active comets such as
LINEAR, higher sensitivity is required. The Combined Array for
Research in Millimeter Astronomy (CARMA), planned to be operational in
2005, will already provide the increase in sensitivity required for a
marginal detection of a comet like LINEAR; with the advent of the
Atacama Large Millimeter Array (ALMA) such objects will become easily
accesible. Needless to say, detailed excitation calculations along the
lines of \citet{bockelee:comethcn}, \citet{crovisier:rovibspectra},
and \citet{biver:hyakutake} will be needed to interpret these future
observations.

\section{Conclusions\label{s:conclusions}}

\begin{enumerate}

\item{We detected HCN 1-0 emission in autocorrelation mode with the
BIMA array at $14\pm 4$~mK~km~s$^{-1}$ in the $143''$ beam, and find a
2$\sigma$ upper limit of $144$~mJy~beam$^{-1}$~km~s$^{-1}$
(203~mK~km~s$^{-1}$) in the $12''\times 9''$ synthesized beam of
combined crosscorrelation data from OVRO and BIMA. Only upper limits
are obtained for HCO$^+$ of $<4$~mK~km~s$^{-1}$ in autocorrelation and
$<80$~mJy~beam$^{-1}$~km~s$^{-1}$ in crosscorrelation. HCN emission
was detected in the three days preceding (2000 July 21.9--24.2) the
breakup of the comet's nucleus around 2000 July 23--24; no emission in
HCN or HCO$^+$ was detected after the breakup.}

\item{Compared to models of HCN excitation in cometary comae, we find
that the relative and absolute integrated intensities in $143''$,
$28''$, and $12''\times 9''$ are consistent with expectations. This
indicates a nuclear source for HCN, and no significant contribution
from a spread-out ($> 1300$~km) distribution of evaporating icy
fragments. Fragments exceeding 1~m in size are thought to be able to
resist evaporation over such distances, and our data limit their
contribution to the volatiles to $\lesssim 50$\% during the days of
2000 July 21--23 immediately preceding the full disruption of comet
LINEAR.}
  
\item{We showed that data obtained at the BIMA and OVRO
interferometers can be successfully combined, creating a `virtual'
CARMA array that increases the signal-to-noise by approximately a
factor of $\sqrt{2}$ due to the increased collecting array. While for
these observations the array configurations were chosen to provide
overlapping, short $uv$ spacings because of atmospheric
phase-stability concerns, virtual `CARMA' observations have the
possibility to offer complementary array configurations increasing
imaging capabilities. These, however, fall short of what a truly
combined array with baselines between the individual BIMA and OVRO
antennas can offer: a combined CARMA array is foreseen to be
operational by 2005. With the arrival of the Atacama Large Millimeter
Array, spatially resolved comet observations at (sub) millimeter
wavelengths will become increasingly accesible.}

\end{enumerate}

\acknowledgments We are indebted to the staffs of the Owens Valley and
Hat Creek Radio Observatories for providing excellent support during
the observations. Wilson Hoffman and Steve Scott are especially
thanked for creating and maintaining the software required to track
fast-moving targets such as this comet. The research of M.~R.~H. in
Berkeley was supported by the Miller Institute for Basic Research in
Science. This work was partially funded by: NASA NAG5-4292, NAG5-4080,
NAG5-8708, and NGT5-0083; NSF AST 96-13998, AST96-13999, AST96-13716,
AST96-15608, and AST99-81363; Taiwanese grants NSC 86-2112-M-003-T and
89-2112-M-003-004; and the Universities of Illinois, Maryland, and
California, Berkeley. We greatly thank the referee,
D.~Bockel\'ee-Morvan, for providing constructive criticism of our paper
and pointing out ways in which to significantly improve our analysis.

\appendix

\section{Combining data from different interferometer
arrays\label{appendix}}

The results presented in this Paper include images that are
constructed based on data obtained simultaneously at two different
millimeter interferometers, BIMA and OVRO. The process of combining
such data contains a number of steps where special care needs to be
taken to ensure that the resulting images correctly represent the data
obtained at the individual arrays. This Appendix describes in detail
the data combination process using the MIRIAD software package, and
presents simulations that validate the procedure (Fig. \ref{f:sim}). 

As a first step, even before the data are taken, one needs to ensure
that both arrays are using the same source position as phase tracking
system, and that the receivers and correlators are configured to
record the same line frequency using the same systemic velocity of the
object at all times. Especially for a moving target like a comet,
these requirements are not trivial when dealing with two separate
observatories.

After the data have been taken, the time variation of the complex
gains (amplitude and phase) has to be calibrated for the two data
sets separately since there is no correlation between the gain
variations at the two sites. It is not necessary that both arrays use
the same objects, usually quasars, as phase and amplitude calibrators.
For the absolute calibration of the flux, however, it can be
advantageous to include the same secondary calibrator (a bright
quasar), the flux of which is derived from contemporaneous
observations of a primary calibrator (planet), preferably with both
arrays. This ensures that there are no offsets in flux calibration
between the arrays. Especially if the arrays have a different response
to spatial scales, such flux offset would affect the resulting
combined image.

Before the calibrated data sets can be combined, they need to be
converted to a common data format (if not already the same). The OVRO
data are usually exported in FITS format, while BIMA data are written
in a format particular to the MIRIAD software package. The MIRIAD task
`FITS' can convert FITS-files to MIRIAD-files. To successfully convert
OVRO data using this task, use `{\tt options=varwt}' so that the
weights of the visibility data written by MMA (the OVRO data
calibration package) are properly interpreted as the reciprocal of the
noise variance. Further processing in the MIRIAD package does not
require any additional conversions: many MIRIAD tasks accept input data
sets that, for example, have different spectral gridding as do our
OVRO and BIMA data.

Figure 3 illustrates the data combination procedure. To test our
combination procedure, this figure also shows the results of combining
data with a relative spatial offset: this results in an image with two
`sources', each with approximately 50\% of the flux because in our
observations the sensitivity of the OVRO and BIMA data was about
equal.



\newpage

\figcaption[Hogerheijde.fig1.ps]{Autocorrelation spectra obtained in
the $143''$ BIMA primary beam. The top row shows single-day averages; the
subsequent rows show the two-day, three-day, and six-day
averages. Emission of HCN 1--0 is detected at $\sim 4\sigma$ averaged
over 2001 July 21.65--24.16, indicated by the grey bar in the
corresponding panel.\label{f:auto}}

\figcaption[Hogerheijde.fig2.ps]{Crosscorrelation images of the
combined BIMA and OVRO data centered on a velocity of 0 km~s$^{-1}$,
at 2$\sigma$ contours (1$\sigma$=42~mJy~beam$^{-1}$~km~s$^{-1}$ for
the single-day averages). The top row shows the single-day averages;
the subsequent rows show the averages over two, three, and six
days. The synthesized beam is indicated in the lower left corner of
each panel.\label{f:cross}}

\figcaption[Hogerheijde.fig3.ps]{The top two panels show the model
image, as observed by OVRO (left) and BIMA (right). After combination,
the `virtual CARMA' result is shown at the lower left. By introducing
an offset in the source position of $+10''$ for the BIMA model and
$-10''$ for the OVRO model, the resulting image in the combined data
(lower right) is split. This latter panel shows that the data
combination procedure in fact preserves the data from both
arrays.\label{f:sim}}

\newpage

\begin{deluxetable}{cccccc}
\tablecaption{Observations\label{t:obs}}
\tablecolumns{6}
\tablehead{
\colhead{OVRO} & \colhead{BIMA} & \colhead{BIMA} &  \colhead{JPL} & &
 \colhead{Calibrator's} \\
\colhead{cross corr.} & \colhead{cross corr.} & \colhead{auto corr.} & 
 \colhead{orbit} & \colhead{Gain} & \colhead{flux}\\
\colhead{UT 2000 July} & \colhead{UT 2000 July} & \colhead{UT 2000 July} & 
 \colhead{solution\#} & \colhead{calibrator\tablenotemark{a}} & 
 \colhead{(Jy)}}
\startdata
18.47--19.15 & \nodata      & \nodata      & 62 & 0716+714 & 3.0 \\
19.47--19.88 & \nodata      & \nodata      & 68 & 0716+714 & 3.0 \\
19.95--19.99 & \nodata      & \nodata      & 68 & 0716+714 & 3.0 \\
20.50--20.93 & 20.45--20.73 & \nodata      & 77 & 0716+714 & 3.0 \\
21.55--22.13 & 21.69--21.82 & 21.85--22.13 & 77 & 0923+392 & 4.5 \\
22.59--23.14 & 22.63--22.92 & 22.95--23.12 & 78 & 0923+392 & 4.5 \\
23.64--24.08 & 23.65--23.92 & 23.95--24.16 & 82 & 0923+392 & 4.5 \\
24.70--25.12 & 24.67--24.92 & 24.95--25.12 & 82 & 0923+392 & 4.5 \\
25.72--26.16 & 25.69--25.94 & 25.99--26.30 & 83 & 0923+391 & 4.5 \\
26.76--27.19 & 26.70--26.89 & 26.92--27.17 & 87 & 1156+295 & 1.2 \\
\enddata
\tablenotetext{a}{Source designation following B1950-based OVRO calibrator 
list, {\tt http://www.ovro.caltech.edu/mm/preobs/guide.html}}
\end{deluxetable}

\begin{deluxetable}{lrrr}
\scriptsize
\tablecaption{HCN 1--0 Intensities: Observations and Models\label{t:results}}
\tablecolumns{4}
\tablehead{
 & \colhead{BIMA} &                    & \colhead{BIMA+OVRO} \\
 & \colhead{autocorr.} & \colhead{IRAM~30-m} & \colhead{crosscorr.} \\
 & \colhead{$143''$}   & \colhead{$28''$}   & \colhead{$12''\times 9''$}
}
\startdata
\cutinhead{Integrated intensities, in mK km~s$^{-1}$}
Observed & $14\pm 4$ & $93\pm 13$\tablenotemark{a} & $<203$ \\
Model 1\tablenotemark{b} & 12.4 & 44.4 & 110.0 \\
Model 2\tablenotemark{c} & 22.7 & 110.0 & 172.8 \\
Model 3\tablenotemark{d} & 5.96 & 24.6 & 89.9 \\
\cutinhead{Integrated intensities, normalized to $28''$ beam}
Observed & $0.15 \pm 0.05$ & $\equiv 1$ & $<2.5$ \\
Model 1 & 0.28 & $\equiv 1$ & 2.15 \\
Model 2 & 0.21 & $\equiv 1$ & 1.57 \\
Model 3 & 0.24 & $\equiv 1$ & 3.65 \\
\enddata
\tablenotetext{a}{Bockel\'ee-Morvan 2000, priv. comm.}
\tablenotetext{b}{$Q$(H$_2$O)=$2\times 10^{29}$~s$^{-1}$, $r_h=1.0$~AU,
 from \citet{crovisier:rovibspectra}, their Fig.~2c.}
\tablenotetext{c}{$Q$(H$_2$O)=$2\times 10^{28}$~s$^{-1}$, $r_h=1.6$~AU,
 from \citet{crovisier:rovibspectra}, their Fig.~3c.}
\tablenotetext{d}{$Q$(H$_2$O)=$1\times 10^{30}$~s$^{-1}$, $r_h=0.8$~AU,
 from \citet{crovisier:rovibspectra}, their Fig.~4c.}
\tablecomments{Observed line intensities and upper limit are averages
over 2000 July 21--23. $Q({\rm HCN})= 3\times 10^{25}$~s$^{-1}$ has
been adopted for the calculations, following \citet{bockelee:linear}.}
\end{deluxetable}

%



\newpage

\begin{figure}
\figurenum{\ref{f:auto}}
\epsscale{1.0}
\plotone{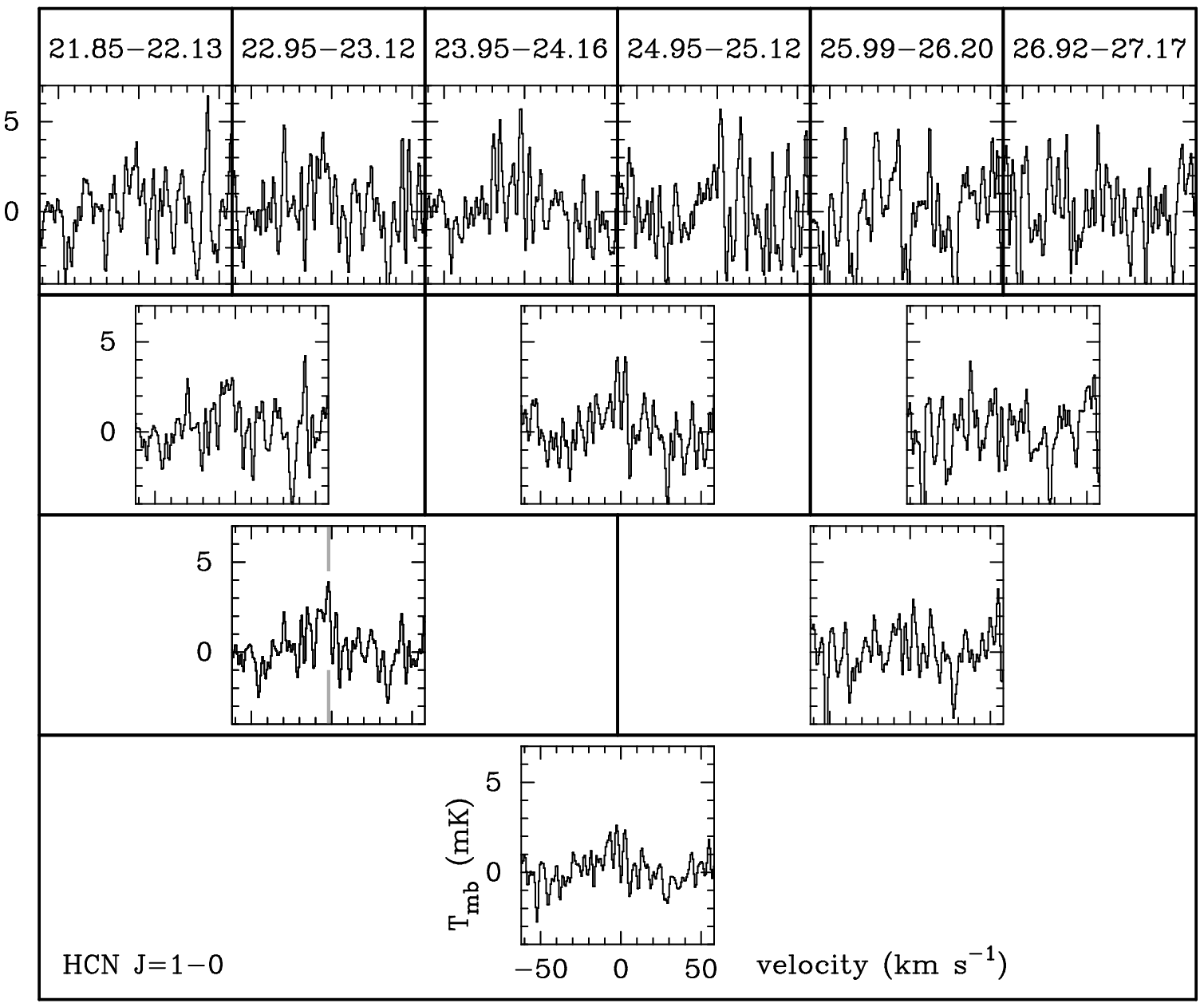}
\caption{}
\end{figure}

\begin{figure}
\figurenum{\ref{f:cross}}
\epsscale{1.0}
\plotone{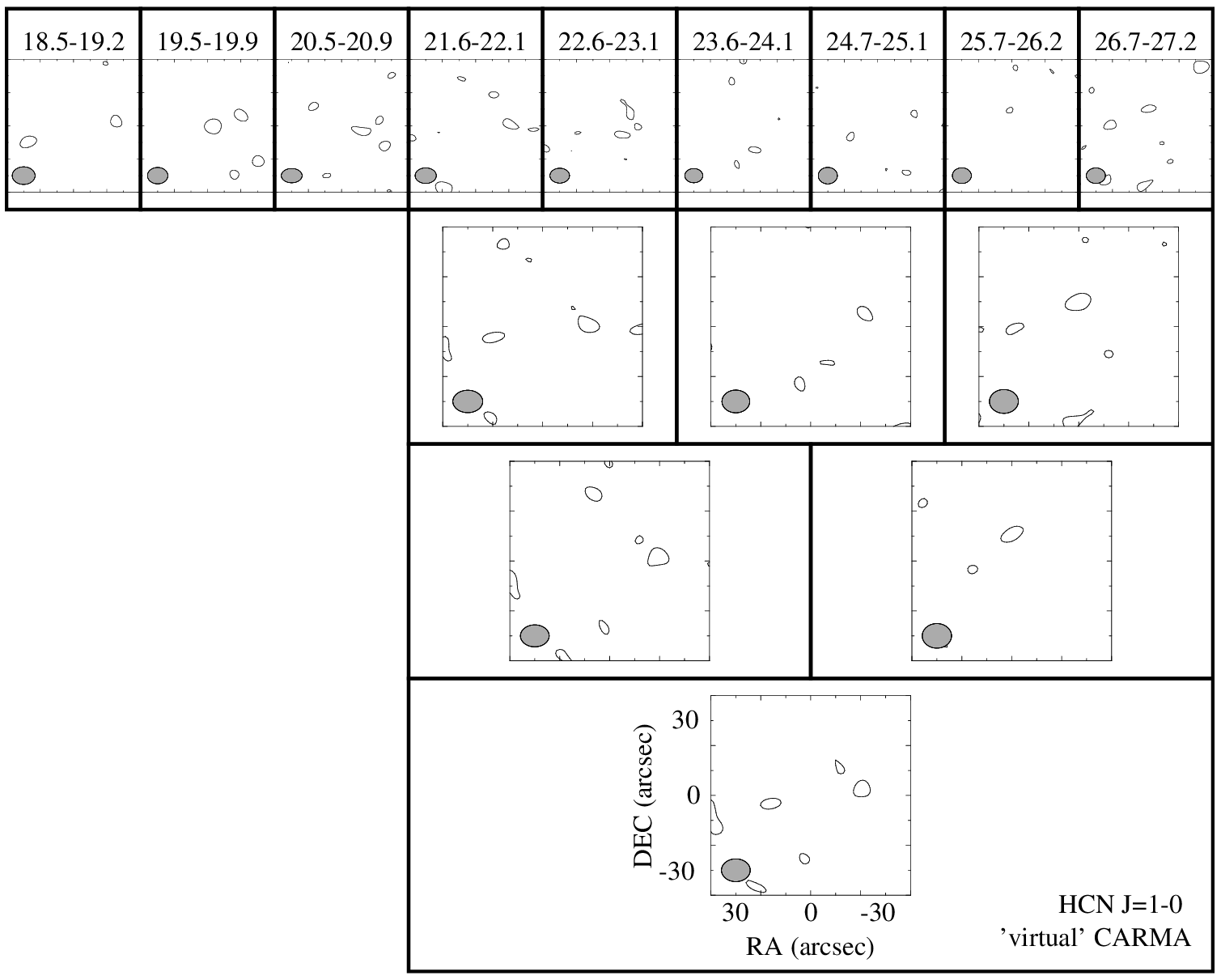}
\caption{}
\end{figure}

\begin{figure}
\figurenum{\ref{f:sim}}
\epsscale{1.0}
\plotone{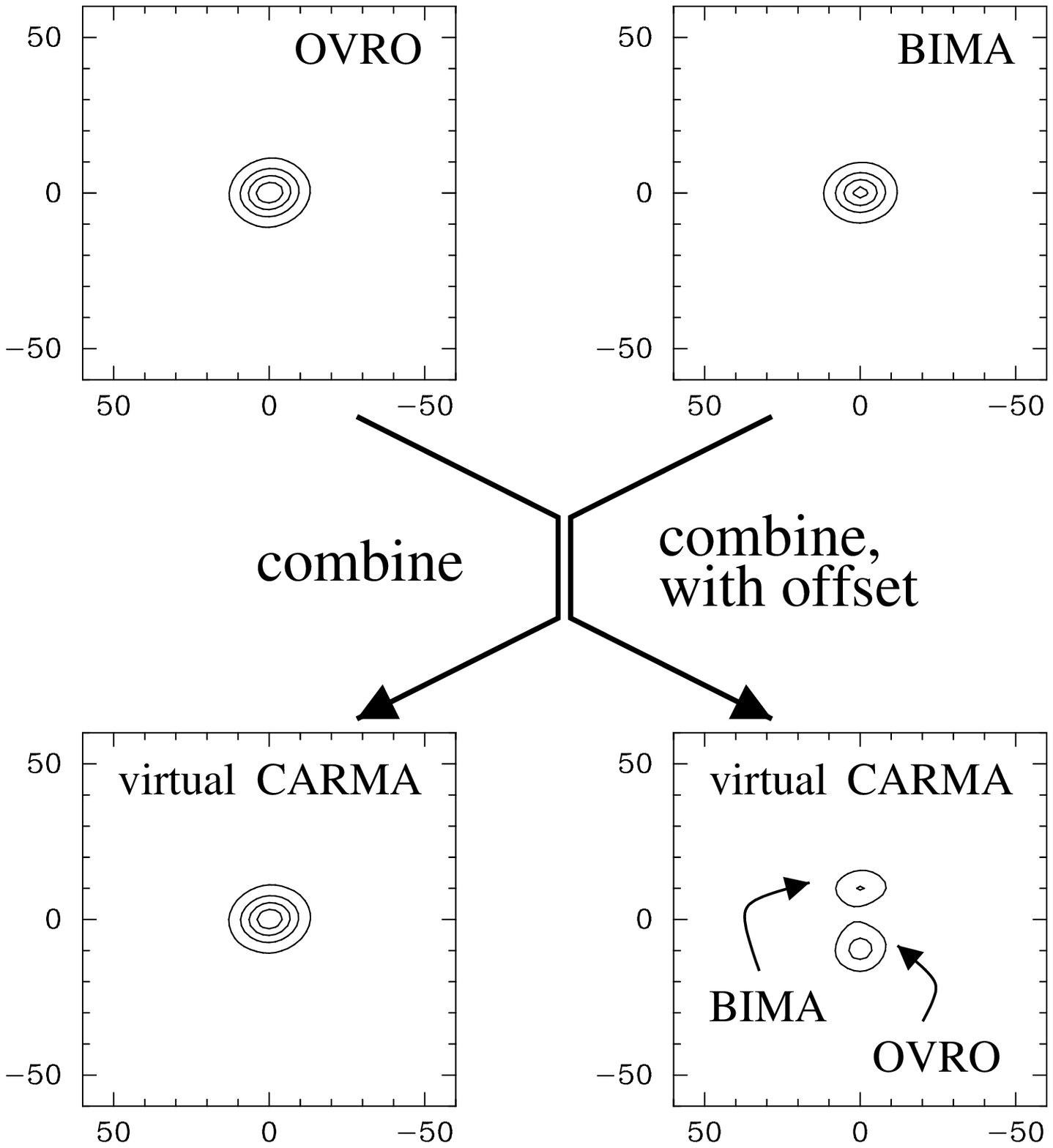}
\caption{}
\end{figure}


\end{document}